\documentstyle[preprint,aps,epsf,amstex]{revtex}

\begin{document}
\draft

%%%%%%%%%%% Titlepage

\begin{titlepage}

\begin{flushright}
\begin{tabular}{l}
NORDITA--98--35--HE\\
SPbU--IP--98--11 \\
hep-ph/9805225
\end{tabular}
\end{flushright}
\vskip0.5cm

\begin{center}
  {\Large \bf
    Integrability of three-particle evolution equations in QCD \\}
  \vskip0.5cm {\large V.M.\ Braun}
%\footnote{On leave of absence from
%  St.\ Petersburg Nuclear Physics Institute, 188350 Gatchina, Russia.}
  \\
  \vskip0.2cm
  NORDITA, Blegdamsvej 17, DK--2100 Copenhagen, Denmark\\
  \vskip0.5cm {\large S.\'{E}.\  Derkachov}
  \\
  \vskip0.2cm
 Department of Mathematics,
 St.-Petersburg Technology Institute, St.-Petersburg, Russia \\
  \vskip0.5cm {\large A.N.\ Manashov}
  \\
  \vskip0.2cm
Department of Theoretical Physics,  Sankt-Petersburg State
University, St.-Petersburg, Russia\\

\vskip1.8cm
  {\large\bf Abstract:\\[10pt]} \parbox[t]{\textwidth}{ 
We show that the Brodsky-Lepage evolution equation for the % leading twist
spin 3/2 baryon distribution amplitude is completely integrable and  reduces
to the three-particle $XXX_{s=-1}$ Heisenberg spin chain. 
Trajectories of the anomalous dimensions are identified and calculated
using the $1/N$ expansion. Extending this result, we prove 
integrability of the evolution equations for 
twist~3 quark-gluon operators in the large $N_c$ limit.
 }

\end{center}

\vskip1.5cm

\begin{center}
%\pacs{
%}
\end{center}

\date{\today}  

\end{titlepage}

\tighten

%\narrowtext

A QCD description of hard exclusive processes 
invokes the concept of hadron distribution amplitudes (DAs)
given by matrix elements of nonlocal light-cone operators bewteen the
vacuum and the hadron state.
It was realized long ago that 
one-loop renormalization group (RG) equations for the leading twist meson 
distributions are diagonalized in the 
conformal basis: The mixing matrix for the corresponding local operators
becomes diagonal and the anomalous dimensions coincide with those familiar
from studies of the deep-inelastic scattering.
The situation with baryon DAs is much more complicated.
In this case one has to consider renormalization of three-quark operators
\cite{nucWF} of the type: 
\begin{equation}
B(a,b,c) =
\varepsilon_{ijk}q^{i}(au)q^{j}(bu)q^{k}(cu).
\label{3q}
\end{equation}
Here $q$ is a quark field  and $u$ is an auxiliary 
light-like vector $u^2=0$; taking 
the leading-twist part and insertion of gauge factors is implied.
Conformal symmetry allows to eliminate all mixing with operators 
including total derivatives but is not sufficient to diagonalize 
the mixing matrix. In the usual expansion of baryon DAs 
in Appell polynomials only the lowest-order term 
(asymptotic DA) is renormalized multiplicatively while in general 
one is left with a nontrivial $(N+1)\times (N+1)$ mixing matrix, $N$ being the 
number of derivatives, and has to diagonalize it explicitly order by order,
see \cite{ren3q,QUA,Ohrndorf}. The analytic structure of the spectrum 
was not known  and further analytic results were not expected.

Main result of this letter is that the 
3-particle Schr\"{o}dinger equation describing the renormalization of
spin 3/2 baryon operators is completely integrable, 
i.e. has an additional
integral of motion. Our result is similar in spirit to the 
recent discovery \cite{Lip,FK} of  
integrability of the system of interacting reggeized gluons in QCD, but 
is obtained in a different physical context. 
In particular, the ERBL evolution 
equation for spin 3/2 baryon operators appears to be mathematically equivalent
to the equation for the odderon trajectory, and the results obtained 
in the latter context \cite{K96,K97} can be adapted to unravel the spectrum of 
baryon operators.

Extending this result, we prove integrability of the RG  equations 
for twist~3 quark-gluon operators 
%\FL
\begin{eqnarray}
S^{\pm}_{\mu}(a,b,c)&=&{\bar q(a u)}
 [iG_{\mu\nu}(b u)\pm
{\tilde G_{\mu\nu}(b u)\gamma_{5}}]u^{\nu}\!
{\not\! u}\, q(c u),
\label{BB} \\
T(a,b,c)&=&{\bar q(a u)} u^{\mu} u^{\nu}
{\sigma_{\mu}}^{\rho}G_{\nu\rho}(b u) \Gamma\,
q(c u), \label{KT} 
\end{eqnarray}
where $ \Gamma=\{I,i\gamma_{5}\} $, in the limit of large number of colors
$N_c$. Such operators give rise to 
twist~3 nucleon parton distributions and have attracted considerable 
interest recently, see e.g.  \cite{ABH,BBKT}. 
Further results will be presented elsewhere \cite{future}. 

\vskip0.5cm

To one-loop 
accuracy, the divergent part of the nonlocal operator $\Phi(a_1,a_2,a_3)$
where $\Phi=B,S^\pm,T$, has the form
$(1/\epsilon)H\Phi(a_i)$, and the explicit expression for
 the integral operator $H$ is known for all cases under
consideration~\cite{QUA,BrBal,BBKT}. 
An arbitrary local operator ${\cal O}$ 
with $N$ covariant derivatives can be represented by 
the associated polynomial in three variables $\psi(a_1,a_2,a_3)$ 
of degree $N$ such that
${\cal O}_\psi = \psi(\partial_a,\partial_b,\partial_c) 
\Phi(a,b,c)_{a,b,c\rightarrow 0}$ 
where $\partial_a=\partial/\partial a$ etc.
In order to find multiplicatively renormalizable local operators one has to 
solve the Schr\"odinger equation for the $\psi$-functions, 
$\widetilde H\psi = {\cal E}\psi$, where $\widetilde H$ is easy 
to find if $H$ is given.

It proves convenient to define the integral transformation \cite{DKM} 
$\psi(a_i)\rightarrow \widehat\psi(z_i)$ by 
\begin{eqnarray}
\widehat\psi(z_i) \equiv \prod_i\int_0^{\infty}dt_i
\,e^{-t_i}t_i^{l_i+s_i-1}\psi(z_{1}t_1,...,z_{3}t_3)
\label{hat}
\end{eqnarray}
where $l_i$ and $s_i$ are the canonical dimension and spin projection of the 
i-th field, respectively:
$l=3/2$ , $s=1/2$ for quarks (antiquarks) and 
$l=2$ , $s=1$ for gluons.
We can reformulate the above eigenvalue problem  in terms
of $\widehat\psi$ functions and it is easy to check that the corresponding 
Hamiltonian  $\widehat H$ coincides with the initial Hamiltonian for 
the nonlocal operator, $\widehat H\equiv H$. As a trivial 
consequence of topology of one-loop Feynman diagrams $H$ has a two-particle
structure, $H=\sum_{i,k}H_{ik}$. Conformal invariance 
implies that the two-particle Hamiltonians $H_{ik}$ 
commute with $SL(2)$ generators  
$ J^{\pm,3} = \sum_{i=1}^{3} J_i^{\pm,3}$, where
\begin{eqnarray}
J^{+}_{i}&=&z_{i}^2\partial_{i}+(l_i+s_i) z_{i},\ \
J^{-}_{i}=-\partial_{i},
\nonumber \\
J^{3}_{i}&=& z_{i}\partial_{i}+{(l_i+s_i)}/{2},
\end{eqnarray}
and are hermitean  with respect to the scalar product
\begin{equation}
\langle \widehat\psi_{1}|\widehat\psi_{2}\rangle =
\psi_{1}(\partial_{1},\ldots,\partial_{3})
\widehat\psi_{2}(z_{1},\ldots,z_{3})\bigl |_{z_i=0}.
\label{sprod}
\end{equation} 
Thus,  the equation
$H\widehat\psi={\cal E}\widehat\psi$ 
decays into the set of eigenvalue problems on the subspaces of functions
with fixed value of $J_{3}$,
$J_{3}\widehat\psi=j_3\widehat\psi$ and annihilated by $J_{-}$,
$J_{-}\widehat\psi=0$.
The advantage of the '$\widehat\psi$-representation' is that 
in this basis the latter condition
is simply shift-invariance \cite{DKM}. Therefore, 
eigenfunctions of two-particle
Hamiltonians are given by simple powers  $\widehat\psi_l=(z_i-z_k)^l$ 
instead of Jacobi polynomials in standard variables \cite{Makeenko}.

The $SL(2)$ invariance imposes stringent restrictions on the form
of two-particle operators, so
that only a few structures are allowed. One such structure corresponds to
the 'vertex correction' involving 
the gluon field from (one of) the covariant derivatives 
(in Feynman gauge):
%\FL
\begin{eqnarray}
H_{12}^{v}\widehat\psi(\underline{z})&=&
-\int_{0}^{1}\!\frac{d\alpha}{\alpha}\Big \{ 
{\bar\alpha}^{l_1+s_1-1}
\left [  
\widehat\psi(z_{12}^\alpha,z_2,z_3)- \widehat\psi(\underline{z})\right]
  \nonumber\\
&&{}
+{\bar\alpha}^{l_2+s_2-1}\!\left [
\widehat\psi(z_1,z_{21}^\alpha,z_3)
-\widehat\psi(\underline{z})
\right ] \!\! \Big \},
\end{eqnarray}
\narrowtext
\noindent
where $\underline{z}\equiv \{z_1,z_2,z_3\}$, 
$z_{ik}^\alpha=z_{i}\bar\alpha+z_{k}\alpha$ and $\bar
\alpha=1-\alpha$. 
Another structure originates from
gluon exchange between quarks (or between a quark and a
gluon):
\begin{equation}
H_{12}^{e}\widehat\psi(\underline{z})=
2\int_{0}^{1}\!\!\!D\underline{\alpha}\,
 \frac{\bar\alpha_{1}^{l_1+s_1}\bar\alpha_{2}^{l_1+s_1}}
{(\bar\alpha_{1}\bar\alpha_{2})^2}
\widehat\psi(z_{12}^{\alpha_{1}},z_{21}^{\alpha_{2}},z_3),
\end{equation}
where $D\underline{\alpha}\equiv
\prod_{i=1}^{3}d\alpha_{i}\,\delta(1-\sum \alpha_{i})$.
 
Due to $SL(2)$ invariance
the two-particle hamiltonians must depend on
the corresponding Casimir
operators ${\rm L}_{ik}\equiv(\vec J_{i}+\vec J_{k})^2=J_{ik}(J_{ik}+1)$
only.  This dependence can be easily reconstructed from the spectrum of
$H_{ik}$. Since the form of the eigenfunctions is known
$ \widehat\psi_{l}=(z_{i}-z_{k})^{l} $,  it is straightforward to derive
\begin{eqnarray}
H_{ik}^{v,(qq)} &=& 2\left[\psi(J_{ik}+1)-\psi(2)\right ],
\\
H_{ik}^{v,(qg)} &=& \psi(J_{ik}+3/2)+\psi(J_{ik}+1/2)-\psi(3)-\psi(2),
\nonumber
\end{eqnarray}
where $\psi(x)\equiv \Gamma'(x)/\Gamma(x)$.
The superscripts $(qq)$ and $(qg)$ indicate quark-quark and 
quark-gluon operators, respectively.
Similarly, we obtain
\begin{eqnarray}
H_{ik}^{e,(qq)} &=& 2 J_{ik}^{-1}(J_{ik}+1)^{-1},
\nonumber\\
H_{ik}^{e,(qg)} &=& 2(J_{ik}+3/2)^{-1}(J_{ik}-1/2)^{-1}.
\end{eqnarray}

We are now in a position to specify RG equations
for the operators in (\ref{3q}--\ref{KT}) explicitly.
One has to distinguish 3-quark operators belonging
to $(3/2,0)$ and $(1,1/2)$ representations, which correspond to DAs
for spin 3/2 and spin 1/2 baryons, respectively. 
We get \cite{remark2}:
%\FL
\begin{mathletters}
\label{def1}
\begin{eqnarray}
H_{3/2}&=&H_{12}^{v,(qq)}+H_{13}^{v,(qq)}+H_{23}^{v,(qq)}, \label{h3q}
      \\
H_{1/2}&=& H_{3/2}-(1/2)H_{12}^{e(qq)}-(1/2)H_{23}^{e,(qq)}.
\label{spin12}
\end{eqnarray}
\end{mathletters} 
Omitting subleading in $N_c$ terms, the quark-antiquark-gluon Hamiltonians
are \cite{ABH,BBKT} 
%\FL
\begin{mathletters}
\label{def2}
\begin{eqnarray}
H_{S^+}&=&H_{12}^{v,(qg)}+H_{23}^{v,(gq)}-H_{12}^{e,(qg)},
\label{sp} \\
H_{S^-}&=&H_{12}^{v,(qg)}+H_{23}^{v,(gq)}-H_{23}^{e,(gq)},
\label{sm}
\\ H_{T}&=&H_{12}^{v,(qg)}+H_{23}^{v,(gq)}-H_{12}^{e,(qg)}-H_{23}^{e,(gq)}.
\label{TT}
\end{eqnarray}
\end{mathletters}
The properly defined anomalous dimensions are given in terms of 
eigenvalues of the above operators including  
color factors and trivial contributions 
of self-energy insertions:
\begin{eqnarray}
  \gamma_{3/2,1/2}(N) &=& (1+1/N_c)\,{\cal E}_{3/2,1/2}(N) +(3/2) \,C_F,
\nonumber\\
  \gamma_{S,T}(N) &=& N_c\, {\cal E}_{S,T}(N) +(7/2)\, N_c
\end{eqnarray}
where $C_F=(N_c^2-1)/(2N_c)$.

The operators $H_{S^\pm}$ are equivalent; hereafter we consider
$H_{S^+}$. 
The $1/N_c^2$ corrections to (\ref{def2}) and RG equations for
3-gluon operators involve additional structures \cite{future} and will 
not be discussed here.

\vskip0.5cm

We have been able to find integrals of motion $Q_i$, $[H_i,Q_i]=0$, 
for all Hamiltonians in question with the exception of $H_{1/2}$. 
Explicit expressions for the conserved charges $Q_i$ present the main
result of this letter: 
\begin{eqnarray}
Q_{3/2} &= & \imath [L_{12},L_{13}]=\
\imath\partial_{1}\partial_{2}\partial_{3}
z_{12}z_{23}z_{31},\label{Q32}\\
Q_{S^{+}}&=&\left \{L_{12},L_{23} \right\}-\frac92 L_{23}-\frac12
L_{12},
\label{QS}\\
Q_{T}&=&\left \{L_{12},L_{23} \right\}-\frac92 L_{12}-\frac92 L_{23},
\label{QT}
\end{eqnarray}
where  $\{\ ,\}$ stands for an anticommutator.
Remarkably, $H_{3/2}$ coincides 
in our representation with the Hamiltonian of the
 $XXX_{s=-1}$ 3-particle Heisenberg spin chain.
Hence the expression in (\ref{Q32})
for the conserved charge $Q_{3/2}$ follows directly from
the corresponding classical result, see also \cite{FK}.

 To check that $ [H_T,Q_T]=0 $, $[H_S,Q_S]=0$ 
 we introduce a complete set of functions \cite{DKM}
$ {\widehat \psi}_{ik}^{n}(z_i,z_k;z_l)$, $ n=0,\ldots,N $ where 
$i,k$ is a fixed pair of indices (say, (1,2)) and $z_l$ is the third 
variable, different from $z_i, z_k$.
The functions
$ {\widehat \psi}_{ik}^{n} $  are obtained by the `hat'-transformation
(\ref{hat}) of the set of polynomials of degree $N$:
\begin{eqnarray}
\psi^{n}_{ik}(z_{i},z_{k};z_{l}) &=& Z_n\!\sum_{m=0}^{N-n}\!K_{m}^n
z_{l}^{m}
(z_{i}+z_{k})^{N-n-m}\psi_{n}(z_{i},z_{k}),
\nonumber\\
\lefteqn{\hspace*{-1.8cm}
K_m^n = \frac{(-1)^{N-n-m}C^{N-n}_{m}}{
(l_l+s_l)_m (2n+l_i+l_k+s_i+s_k)_{N-n-m}},}
\label{fundef}
\end{eqnarray}
where
$C^{n}_{k}$ is the binomial coefficient and
$(a)_k \equiv \Gamma(a+k)/\Gamma(a)$.
 $\psi_{n}(z_i,z_k)$ is defined such that
$\widehat\psi_{n}(z_{i},z_{k})=(z_{i}-z_{k})^n$ is an eigenfunction
of $H_{ik}$ and the factor
$Z_n$ is chosen by requiring that ${\widehat \psi}_{ik}^{n} $
has unit norm.

The functions
${\widehat\psi}^{n}_{ik} $ are shift invariant and mutually
orthogonal with respect to the scalar product (\ref{sprod}).
It is easy to check that 
$ J_{3}{\widehat\psi}^{n}_{ik}=(N+7/2){\widehat\psi}^{n}_{ik} $.
The operator
$ L_{ik} $ is diagonal in the basis
${\widehat\psi}^{n}_{ik} $ while the other two Casimir operators
 are three-diagonal 
($\langle n| L| n'\rangle\neq0 $ for $ |n-n'|\leq 1 $ only).

Consider matrix elements of the commutator
$$ A = [Q_{T}, H_{T}^{(12)}] = \{L_{12}-9/4,[H_{T}^{(12)},L_{23}]\}$$
where $H_{T}^{(12)}=H_{12}^{v,(qg)}-H_{12}^{e,(qg)} $, sandwitched 
between the  
$ \widehat\psi_{12}^{n} $ 'states'.
Since $ L_{23} $ is three-diagonal in this basis, the only
nonzero  elements are
$ A_{n,n+1} $ and $ A_{n+1,n} $. Due to antihermiticity of
$ A $ it is sufficient to consider
$ A_{n,n+1} $:
$$
A_{n,n+1}=(L_{23})_{n,n+1}({\cal E}_{n}-{\cal E}_{n+1})
({\cal L}_{n}+{\cal L}_{n+1}-9/2),
$$
where ${\cal E}_{n}$ and $ {\cal L}_{n} $ are the eigenvalues of 
the operators $ H_{T}^{(12)}$ and $L_{12} $, respectively. 
Using explicit expressions
for $ H_{T}^{(12)}, L_{12} $ it is easy to derive
that $A_{n,n+1}= 2(L_{23})_{n,n+1}({\cal
L}_{n}-{\cal L}_{n+1})$, that is in operator form
\begin{equation}
[Q_{T},H^{(12)}_{T}]=2[L_{12},L_{23}].
\end{equation}
Similarly, we obtain that
$ [Q_{T},H_{T}^{(23)}]=2[L_{23},L_{12}] $ and, consequently,
$ [Q_{T},H_{T}]=0 $.  The proof for 
$ S^+ $ operator is analogous.

\vskip0.5cm

Once conserved charges are known, one can consider the eigenvalue 
problem for these charges instead of the Hamiltonians, which is simpler.
For the Heisenberg spin chain, a detailed study
exists due to Korchemsky \cite{K96,K97}. 
The spectrum of $Q_{3/2}$ is shown in Fig.~1a.
For generic integer $N$ there exist $N+1$ eigenvalues which come
in pairs $\pm q$. Note that for even $N$ $Q_{3/2}$ has zero 
eigenvalue $q=0$.
The corresponding value of energy  can be calculated exactly:
\begin{equation}
 {\cal E}_{3/2}(N,q=0) = 4\psi(N+3)+4\gamma_E-6,
\label{lowbound}
\end{equation} 
see the dotted curve in Fig.~1b.
Eigenvalues of $Q_{3/2}$ lie on trajectories which were found
 in \cite{K96} using a 'semiclassical' expansion in the 
parameter $h=N+3$:
\begin{eqnarray}
 q(N,k)/h^3 &=& \sum_{m} q^{(m)}(k)/h^m,
\nonumber\\
 q^{(0)} = 1/\sqrt{27},&\quad&
 q^{(1)}(k) = -(k+1)/\sqrt{3},\ldots
\label{qexpand}
\end{eqnarray}
The $q^{(m)}(k)$ are polynomials of degree $m$; first eight of them
are given in Eq.~(5.14) in Ref.~\cite{K96}.
$k$ is a nonnegative integer which numerates the trajectory. 
The asymptotic expansion in (\ref{qexpand})
is valid for $q>0$ only and the analytic continuation of the trajectory
to $q<0$ can be obtained by using symmetry properties of the solutions 
\cite{K96}:
$$
 q(N,k) \stackrel{q<0}{\longrightarrow} -\,q(N,N-k)
$$ 
Two examples of the trajectories  with $k=2$ and $k=7$ are shown in 
Fig.\ 1a together with exact eigenvalues (crosses) calculated numerically.
%Exact eigenvalues obtained by numerical solution are shown by crosses.
%The agreement is  good and 
Note that the two asymptotic expansions ---
for positive and negative $q$ --- match reasonably well.
Explicit analytic formulas for the trajectories in the $q\to 0$
region are available from \cite{K97}.

The low-lying eigenvalues of $q(N,k)$
can be calculated to $O(1/N)$ accuracy from the equation \cite{future}
\begin{eqnarray}
 (q/N^2)\ln N - arg[\Gamma(1+iq/N^2)]
 &=&\frac{\pi}{6}\big(N-2k\big)
\label{sq3}
\end{eqnarray}
which is valid for $k-N/2\ll \ln N$.
%where $\bar q = q/N^2$, 
%$\tilde k \in {\Bbb Z}$, 
%$\tilde k $ is an integer, 
%$\delta_N=0$ for even $N$ and
%$\delta_N=1$ for odd $N$
%and $\tilde k = (N-\delta_N)/2-k$.
The lowest value of $|q|$ for odd N 
is thus of order
\begin{eqnarray}
q/N^2 &=& \pm\frac\pi6 (\ln N+\gamma_{E})^{-1} +O(1/(\ln N+\gamma_E)^{4}).
\label{qlow}
\end{eqnarray}

The spectrum  of  $H_{3/2}$ is shown in Fig.~1b.
Exact eigenvalues obtained by explicit diagonalization of the mixing 
matrix are shown by crosses. Since ${\cal E}_{3/2}(q) ={\cal E}_{3/2}(-q)$
all eigenvalues except for the ones for $q=0$ are double-degenerate.
The energy eigenvalues lie on trajectories corresponding to the 
trajectories for $q$  in Fig.~1a, and, similar to the latter,
can be calculated using a `semiclassical' expansion \cite{K96}
\begin{eqnarray}
 {\cal E}_{3/2}(N,k) &=& \varepsilon^{(0)}-
                     \sum_{m=1}^\infty \varepsilon^{(m)}(k)/h^m,
\nonumber\\
 \varepsilon^{(0)} &=& 6 \ln(N+3)+6\gamma_E-6-3 \ln 3, \ldots
\label{eexpand}
\end{eqnarray}     
The polynomials $\varepsilon^{(m)}(k)$ are given in Eq.~(6.5) of 
Ref.~\cite{K96} up to $m=7$. 
The trajectories corresponding to $k=2$ and $k=7$ are shown 
in Fig.~1b by broken lines, whereas the solid curves correspond
to the asymptotic expansion in (\ref{eexpand}) \cite{remark1}. Note that the 
expansion diverges close to the 'deflection points'
which occur at even integer $N$ and with the energy 
given by Eq.~(\ref{lowbound}). Convergence of the $1/h$ 
expansion is somewhat worse for the energy compared to  the 
conserved charge $q$, but it can be improved systematically.  
Alternatively, one can derive analytic approximations 
for the conserved charge $q(N,k)$ and  
${\cal E}_{3/2}(q)$ applicable in the $q\rightarrow 0$ region, see 
\cite{K97,future}.

Using (\ref{qlow}) one can derive an estimate for the lowest energy
eigenvalue for odd $N$:
\begin{equation}
{\cal E}_{3/2}(N) = 4 \ln N - 6 +4\gamma_E 
 + \frac{\zeta(3)}{18 \ln^2N}.
\end{equation}
Numerically the difference to Eq.(\ref{lowbound}) is very small, 
compare exact eigenvalues with the dotted curve in Fig.~1b, and is 
probably irrelevant for phenomenological applications. One has to bear
in mind, however, that an approximation of taking into account 
operators with the lowest anomalous dimension only for each $N$ is 
theoretically inconsistent since they belong to different trajectories. 

%%%%%%%%%%%%%%%%%
\vskip0.5cm
%%%%%%%%%%%%%%%%%

The anomalous dimensions of quark-gluon operators 
(\ref{QS}), (\ref{QT}) can be studied along similar lines \cite{future}. 
To the $O(1/N)$ accuracy the spectrum of 
low-lying energy eigenvalues is given in terms of eigenvalues of the 
corresponding integrals of motion as
\begin{eqnarray}
{\cal E}(\nu)&=&2 \ln N+4\gamma_E - 5 + 2{\cal R}e
\left [ \psi(3/2+i\nu) \right]
\end{eqnarray} 
where $2\nu_{S,T}^2 = q_{S,T}-3/2$, and 
quantization conditions for the effective charges read, 
to the same accuracy 
\begin{eqnarray}
&&\nu_T\ln N+\Phi_1(\nu)-\Phi_3(\nu)=\frac{\pi n}{2},
\nonumber\\
&&\nu_S\ln N+\frac12(\Phi_1(\nu)+\Phi_2(\nu))-\Phi_3(\nu)=\frac{\pi n}{2}
\end{eqnarray}
where $n=1,2,\ldots$ and
\begin{eqnarray}
\Phi_1(\nu)&=&(1/2)\,arg\,[{}_2\!F_1^2(3/2+i\nu,-3/2+i\nu,1+2i\nu,1)],
\nonumber\\
\Phi_2(\nu)&=&arg\,[{}_2\!F_1(1/2+i\nu,-1/2+i\nu,1+2i\nu,1)],
\nonumber\\
\Phi_3(\nu)&=& arg\,[ \Gamma(3/2+i\nu)].
\end{eqnarray}
These formulas are not applicable to the exact solutions with minimum 
anomalous dimensions, found in Refs.~\cite{ABH,BBKT}, which correspond
to imaginary $\nu$ and have to be treated separately. We can show
that these special solutions are separated from the rest of the spectrum
by a finite gap. A detailed study is in progress. 

To summarize, we have shown that a few important three-particle 
evolution equations in QCD are exactly integrable, that is they possess 
nontrivial integrals of motion. This allows for a fairly complete 
description of the spectrum of anomalous dimension of baryon 
operators with spin 3/2, and similar techniques can be developed 
for other cases as well. The eigenfunctions can also be studied 
\cite{K97,future}. We believe  that the approach based on integrability 
may find many phenomenological applications to studies of 
higher-twist parton distributions in QCD. 

%%%%%%%%%%%%%%%%%
\vskip0.5cm
%%%%%%%%%%%%%%%%%
\paragraph*{Acknowledgements.}
%\acknowledgments
V.B. is grateful to G.\ Korchemsky for many interesting discussions
and critical reading of the manuscript.
The work by S.D. and A.M. was supported by the RFFR Grant 97--01--01152.

%%%%%%%%%%%%%%%%%%%%%%%%%
% FIGURE ENERGY SPECTRUM
%%%%%%%%%%%%%%%%%%%%%%%%%
\begin{figure}
\centerline{\epsfxsize11.0cm\epsffile{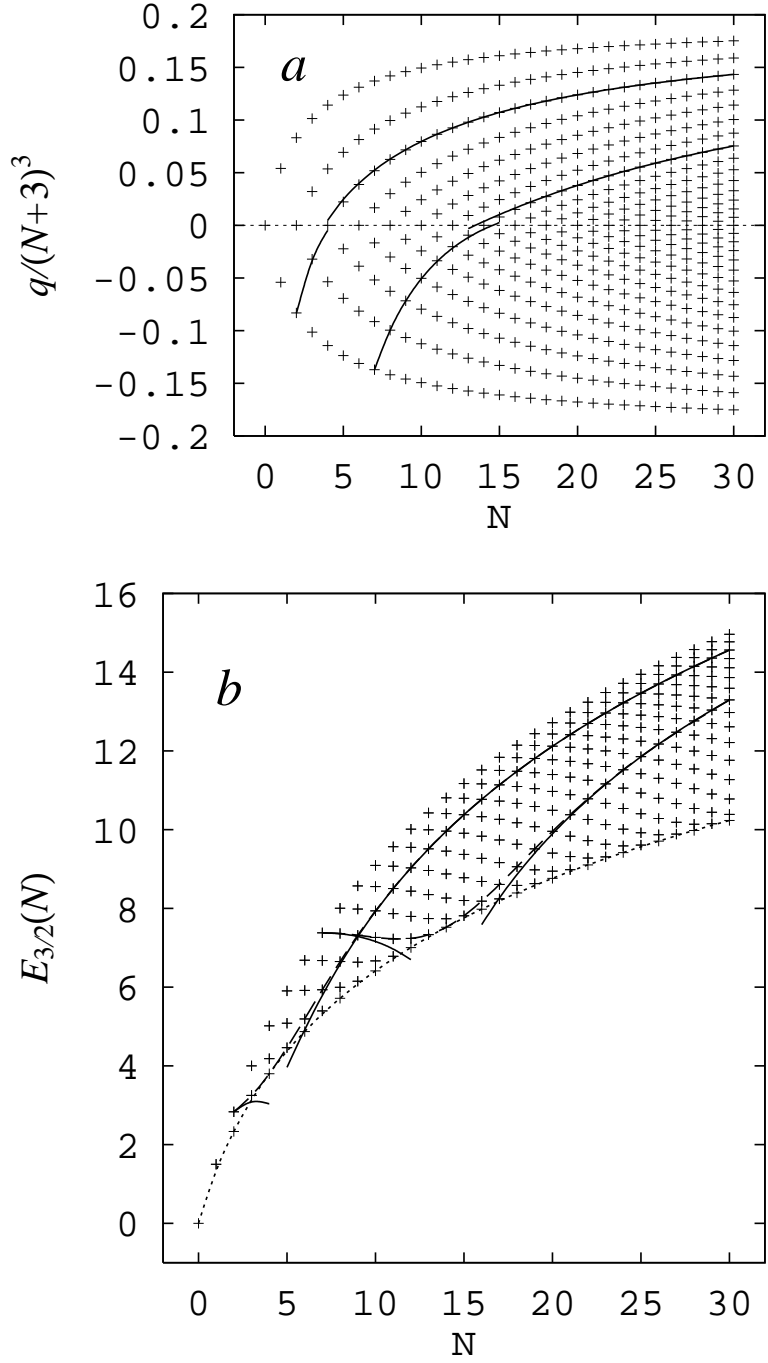}}
\caption[]{The spectrum of eigenvalues for the conserved charge $Q_{3/2}$ (a)
and for the spin 3/2 Hamiltonian $H_{3/2}$ (b), see text.  
}
\end{figure}

\begin{references}
\bibitem{remark2}
We assume here that all three quarks are different; otherwise the
terms $H_{ik}^{e(qq)}$ in (\ref{spin12}) have to be multiplied by 
the appropriate symmetrization factors, cf.~\cite{QUA}.
\end{references}
\end{document}